\documentclass[prb,twocolumn,floatfix]{revtex4}
\usepackage{amssymb}
\usepackage{graphicx}

\def\nbZ{{
\mathchoice {\hbox{$\sf\textstyle Z\kern-0.4em Z$}}
{\hbox{$\sf\scriptstyle Z\kern-0.3em Z$}}
{\hbox{$\sf\scriptscriptstyle Z\kern-0.2em Z$}} }}
\begin{document}
\title{Theory of huge tunneling magnetoresistance in graphene}
\author{Feng Zhai}
\email{fengzhai@dlut.edu.cn} \affiliation{School of Physics and
Optoelectronic Technology and College of Advanced Science and
Technology, Dalian University of Technology, Dalian 116024,
People's Republic of China}
\author{Kai Chang}
\email{kchang@red.semi.ac.cn} \affiliation{NLSM, Institute of
Semiconductors, Chinese Academy of Sciences, P.O. Box 912, Beijing
100083, People's Republic of China}

\begin{abstract}
We investigate theoretically the spin-independent tunneling
magnetoresistance effect in a graphene monolayer modulated by two parallel
ferromagnets deposited on a dielectric layer. For the parallel magnetization
configuration, Klein tunneling can be observed in the transmission spectrum,
but at specific oblique incident angles. For the antiparallel magnetization
configuration, the transmission can be blocked by the magnetic-electric
barrier provided by the ferromagnets. Such a transmission discrepancy
results in a tremendous magnetoresistance ratio and can be tuned by the
inclusion of an electric barrier.
\end{abstract}

\pacs{73.23.-b, 03.65.Pm, 73.43.Cd, 75.70.Ak} \maketitle

Recent experiments have demonstrated the stability of graphene (a
single atomic layer of graphite) and the feasibility of
controlling its electrical properties by local gate
voltages,\cite{Graphene fabrication1, Graphene
fabrication2,Graphene fabrication3,half QHE,gate control1,gate
control2} opening a promising way to explore carbon-based
nanoelectronics. In graphene, the energy spectrum of carriers
consists of two valleys labeled by two inequivalent points
(referred to as $K$ and $K^{\prime }$) at the edges of the
hexagonal Brillouin zone. In each valley, the energy dispersion
relation is approximately linear near the points where the
electron and hole bands touch. Such a peculiar band structure
results in many interesting phenomena, including the half-integer
quantum Hall effect\cite{Graphene
fabrication2,Graphene fabrication3,half QHE} and minimum conductivity.\cite%
{Graphene fabrication2,Graphene fabrication3} Further, Dirac-like fermions
in graphene can transmit through high and wide electrostatic barriers almost
perfectly, in particular for normal incidence.\cite{Klein tunneling1,Klein
tunneling2,Klein tunneling3} Such a phenomenon, known as Klein tunneling,
leads to a poor rectification effect in graphene p-n junctions\cite{gate
control2} and thus may limit the performance of graphene-based electronic
devices.

Very recently, inhomogeneous magnetic fields on the nanometer scale have
been suggested to confine massless two-dimensional (2D) Dirac electrons,\cite%
{magnetic confinement} providing another clue to the manipulation
of electrons in graphene. For conventional semiconductor
two-dimensional electron gas systems, the patterned local magnetic
fields define various
magnetic nanostructures ranging from magnetic barriers and wells\cite%
{magnetic barrier1} to magnetic dots and antidots.\cite{magnetic barrier2} A
great deal of experimental and theoretical works have been devoted to
understand physical properties of Schr\"{o}dinger fermions in these systems.
The effects of nonuniform magnetic field modulations on 2D Dirac-Weyl
fermions, however, has not been investigated as thoroughly, especially for
the Klein tunneling under inhomogeneous magnetic field. In this work we
explore ballistic transport features of graphene under the modulations of
both local magnetic fields and local electrostatic barriers generated by two
parallel ferromagnetic stripes. A remarkable tunneling magnetoresistance
(TMR) effect is predicted and its physical mechanism is explained.

\begin{figure}[tb]
\includegraphics[width=8.0cm,height=8.0cm]{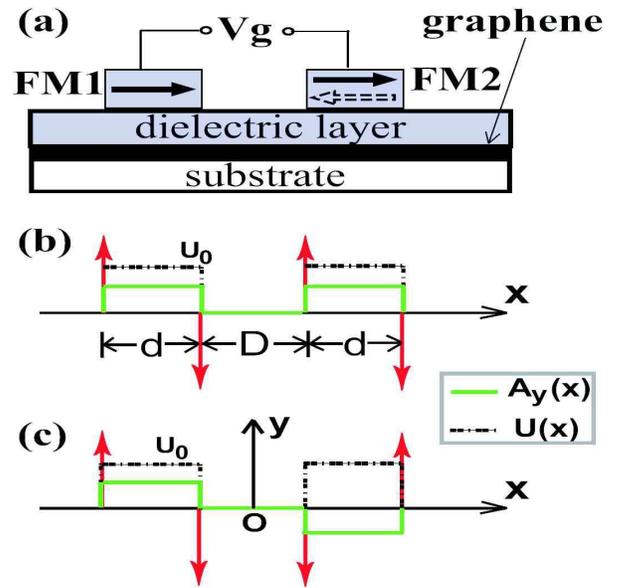}
\caption{(Color online) (a) Schematic illustration of the
considered two-dimensional electron system modulated by two FM
stripes deposited on top of the graphene plane. Each FM stripe has
a rectangular cross section and a magnetization directed along the
current direction (the $x$ axis). The gate voltage $V_{g}$ applied
on both FM stripes provides an electrostatic double barrier in the
underneath graphene plane. (b) Simplified profiles of the magnetic
barrier for the P alignment (spikelike lines), the corresponding
vector potential $A_{y}(x)$ (solid line), and the electrostatic potential $%
U(x)$ (dashed line). (c) The same as in (b) but for the AP
alignment.} \label{system}
\end{figure}
The system under consideration is a single-layer graphene sheet covered by a
thin dielectric layer,\cite{gate control1,gate control2} as sketched in Fig.~%
\ref{system}(a). Two parallel ferromagnetic metal (FM) stripes are
deposited on top of the insulating layer to influence locally the
motion of Dirac electrons in the graphene ($x,y$) plane. Both FM
stripes have a width $d$ and a magnetization in parallel or in
antiparallel to the current direction (the $x$ axis). Their fringe
fields thus provide a perpendicular magnetic modulation $B_{z}$,
which is assumed to be homogeneous in the $y$ direction and only
varies along the $x$ axis. A suitable external in-plane magnetic
field can change the relative orientation of the two
magnetizations which are antiparallel at zero field. At the limit
of a small distance between the graphene plane and the
ferromagnets, the magnetic barrier can be approximated by several
delta functions, i.e., $B_{z}(x)=Bl_{B_{0}}\{[\delta
(x+L/2)-\delta (x+D/2)]+\gamma \lbrack \delta (x-D/2)-\delta
(x-L/2)]\}$.
Here, $B$ gives the strength of the local magnetic field, $l_{B_{0}}=\sqrt{%
\hbar /eB_{0}}$ is the magnetic length for an estimated magnetic field $%
B_{0} $, $\gamma $ represents the magnetization configuration
[$\pm 1$ or parallel (P)/antiparallel (AP)], $D$ is the distance
between the two FM stripes, and $L=2d+D$ is the total length of
the structure along the transport direction. The model
magnetic field configurations for $\gamma =\pm 1$ are depicted in Figs.~\ref%
{system}(b) and \ref{system}(c), respectively. Further, when a negative gate
voltage is applied to both FM stripes, a tunable electrostatic double
barrier potential $U(x)$ arises in the graphene layer. A square shape with
height $U_{0}$ can be taken for the electric potential created by either
gate. Accordingly, the simplified electrostatic barrier has the form, $%
U(x)=U_{0}[\Theta (x+L/2)\Theta (-D/2-x)+\Theta (x-D/2)\Theta (L/2-x)]$,
where $\Theta (x)$ is the Heaviside step function.

For such a system, the low-energy excitations in the vicinity of the $K$
point can be described by the following Dirac equation
\begin{equation}
\left[ \upsilon _{_{F}}\mathbf{\sigma }\cdot (\mathbf{p}+e\mathbf{A}%
)+U\sigma _{0}\right] \Psi =E\Psi \text{,}  \label{Dirac}
\end{equation}%
where $\upsilon _{_{F}}\thickapprox 0.86\times 10^{6}$ m/s is the Fermi
velocity of the system, $\sigma _{x}$, $\sigma _{y}$, and $\sigma _{z}$ are
three isospin Pauli matrices, $\mathbf{p}=(p_{x},p_{y})$ is the electron
momentum, $\mathbf{A}$ is the vector potential which in the Landau gauge has
the form $\mathbf{A=}(0,A_{y}(x),0)$, and $\sigma _{0}$ is the $2\times 2$
unit matrix. Since the Dirac Hamiltonian of graphene is valley degenerate,
it is enough to consider the $K$ point.\cite{magnetic confinement} For
convenience we express all quantities in dimensionless units by means of two
characteristic parameters, i.e., the magnetic length $l_{B_{0}}$ and the
energy $E_{0}=\hbar \upsilon _{_{F}}/l_{B_{0}}$. For a realistic value $%
B_{0}=0.1$ T, we have $l_{B_{0}}=811$ \AA\ and $E_{0}=$ $7.0$ meV.

Since the system is homogeneous along the $y$ direction, the transverse wave
vector $k_{y}$ is conserved. At each region with a constant vector potential
$A_{y}$ and electrostatic potential $U$, the solution of Eq.~(\ref{Dirac})
for a given incident energy $E$ can be written as
\begin{equation}
\Psi =e^{ik_{y}y}\left[ C_{+}e^{ik_{x}x}\left(
\begin{array}{c}
1 \\
\frac{k_{x}+iq}{E-U}%
\end{array}%
\right) +C_{-}e^{-ik_{x}x}\left(
\begin{array}{c}
1 \\
\frac{-k_{x}+iq}{E-U}%
\end{array}%
\right) \right] \text{.}  \label{solution}
\end{equation}%
Here $q=k_{y}+A_{y}$, and $k_{x}$ is the longitudinal wave vector satisfying
\begin{equation}
k_{x}^{2}+(k_{y}+A_{y})^{2}=(E-U)^{2}\text{.}  \label{kx}
\end{equation}%
The sign of $k_{x}$ is chosen in such a way that the corresponding
eigenstate is either propagating or evanescent in the forward direction. The
coefficients $C_{+}$ and $C_{-}$ are determined by the requirement of wave
function continuity and the scattering boundary conditions. The scattering
matrix method\cite{SM} is adopted to obtain these coefficients and the
transmission probability $T=T_{P/AP}(E,k_{y})$ for the P/AP configuration.
The latter depends on the incident energy $E$ and the transverse wave vector
$k_{y}$. The ballistic conductance at zero temperature is calculated from
\begin{eqnarray}
G_{P/AP}(E_{F}) &=&\frac{4e^{2}}{h}\int_{-E_{F}}^{E_{F}}T_{P/AP}(E_{F},k_{y})%
\frac{dk_{y}}{2\pi /L_{y}}  \nonumber \\
&=&G_{0}\int_{-\pi /2}^{\pi /2}T_{P/AP}(E_{F},E_{F}\sin \theta )\cos \theta
d\theta \text{,}  \nonumber \\
&&  \label{conductance}
\end{eqnarray}%
where $L_{y}\gg L$ is the sample size along the $y$ direction, $\theta $ is
the incident angle relative to the $x$ direction, and $%
G_{0}=2e^{2}E_{F}L_{y}/(\pi h)$ is taken as the conductance unit.

\begin{figure}[tb]
\includegraphics[width=8.0cm,height=8.0cm]{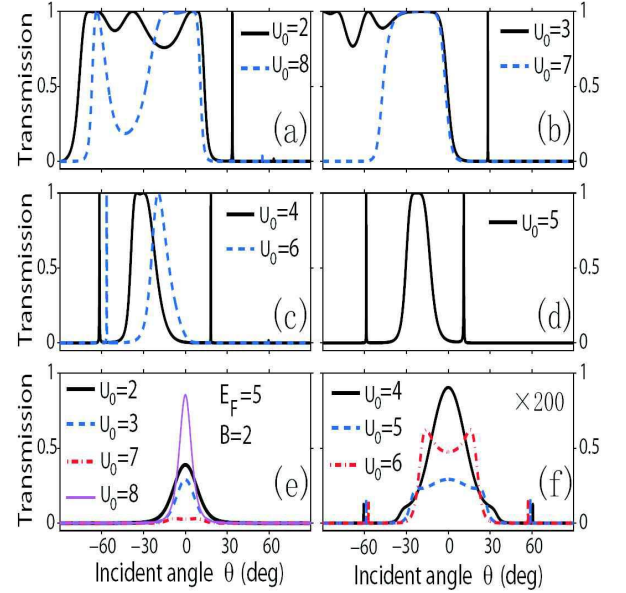}
\caption{(Color online) Transmission as a function of incident angle for
electrons traversing the considered structure (depicted in Fig.~\protect\ref%
{system}) with (a)-(d) parallel or (e)-(f) antiparallel
magnetization configuration. Device parameters used in the calculations are $d=D=1$ and $%
B=2$. The incident energy is fixed at $E=5$. Note that all curves in (f) are
scaled by a factor $200$.}
\label{transmission}
\end{figure}

The proposed device relies on the interplay between the Klein
tunneling and the wave vector filtering provided by local magnetic
fields. To obtain a quantitative understanding of this interplay,
Fig.~\ref{transmission} plots
the transmission probability calculated as a function of the incident angle $%
\theta $ for both the P and AP magnetization configurations. In our
calculations the structure parameters of the magnetic barrier are set at $%
d=D=1$ and $B=2$. The incident energy is fixed at $E=5$ and the electric
barrier height $U_{0}$ is taken as $2,3,4,5,6,7,8$ for different curves.

For the magnetic barrier with P alignment, the transmission spectrum
demonstrates obvious angular anisotropy [see Figs.~\ref{transmission}(a)-\ref%
{transmission}(d)]. The reflection at normal incidence is finite and is
almost complete at suitable electric barrier heights. Instead, perfect
transmission appears at some oblique incidences. For example, in the special
case $E=U_{0}$, the transmission peak with a finite width locates at $%
k_{y}=-A_{y}$ [see Fig.~\ref{transmission}(d)]. In comparison with the case
of pure electric barriers,\cite{Klein tunneling3} we can see that the
magnetic barrier changes the incident direction at which the Klein tunneling
occurs. The transmission is remarkable in a wide region of negative $\theta$
and is blocked by the magnetic barrier when the incident angle exceeds a
critical value $\theta _{+}(U_{0})$ or is below another critical value $%
\theta _{-}(U_{0})$. This can be understood as follows. From
Eq.~(\ref{kx}) we know that evanescent states appear in the
magnetic barrier regions when the magnetic vector potential (here
$A_{y}=B$) and electrostatic barrier satisfy $\left\vert
k_{y}+B\right\vert >\left\vert E-U_{0}\right\vert $. The
transmission is generally weak as the decaying length of the
evanescent states is shorter than the barrier width. In the
transmission forbidden region, there may exist one or two
line-shaped peaks with unity values, as a result of resonant
tunneling through the symmetric double barrier structure. The
applied electric barrier significantly alters the positions of the
transmission peaks. We can also observe a large difference between
the transmission curves for the barrier height $U_{0}<E$ and
$U_{0}^{\prime }=2E-U_{0}$. Such a difference arises from the fact
that the carrier states for the two cases are not completely
complementary.

We next examine the transmission characteristics for the AP alignment, which
is shown in Figs.~\ref{transmission}(e) and \ref{transmission}(f). In this
configuration the magnetic vector potential is antisymmetric about the
central line $x=0$ [see Fig.~\ref{system}(c)]. The Dirac Hamiltonian
possesses a symmetry associated with the operation $\hat{T}\hat{R}_{x}\hat{%
\sigma}_{y}$, where $\hat{T}$ is the time reversal operator and $\hat{R}_{x}$
is the reflection operator $x\rightarrow -x$. This symmetry implies the
invariance of the transmission with respect to the replacement $%
k_{y}\rightarrow -k_{y}$, as seen in Figs.~\ref{transmission}(e)
and \ref{transmission}(f). For large $\left\vert
E-U_{0}\right\vert $ the transmission decays monotonically as the
incident angle increases from zero [see
Fig.~\ref{transmission}(e)]. Since the carrier states in the two
magnetic barriers are not identical, perfect transmission can not
be achieved (except for the case of normal incidence). Note that
for the AP configuration and a given wave vector $k_{y}\geqslant
0$, the presence of
evanescent states in the first magnetic barrier only requires $%
k_{y}>\left\vert E-U_{0}\right\vert -B$. When $\left\vert
E-U_{0}\right\vert <B$ this condition is met for all incident
directions and the transmission can be strongly suppressed, as
shown in Fig.~\ref{transmission}(f). Within this parameter regime,
the transmission exhibits a nonmonotonic variation with the
positive incident angle. Furthermore, the maximal transmission for
the AP alignment can be 2 orders of magnitude lower than that for
the P alignment [Figs.~\ref{transmission}(c) and
\ref{transmission}(d)].

\begin{figure}[tb]
\includegraphics[width=8.0cm,height=6.0cm]{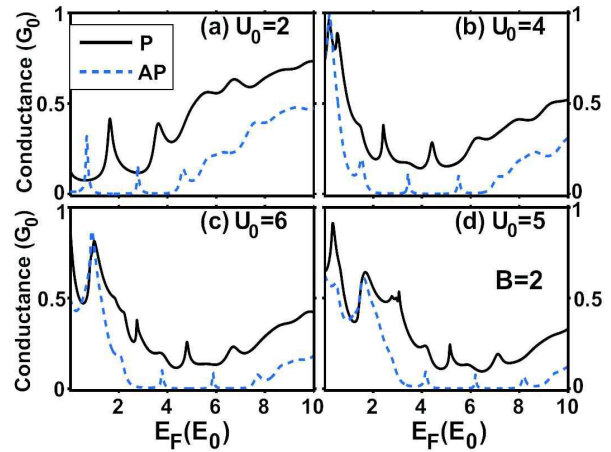}
\caption{(Color online) Conductance as a function of Fermi energy for
electrons traversing the considered structure with a parallel (solid line)
or antiparallel (dashed line) magnetization configuration. Device parameters
used in the calculations are $d=D=1$ and $B=2$.}
\label{Conductance}
\end{figure}

\begin{figure}[tb]
\includegraphics[width=8.0cm,height=6.0cm]{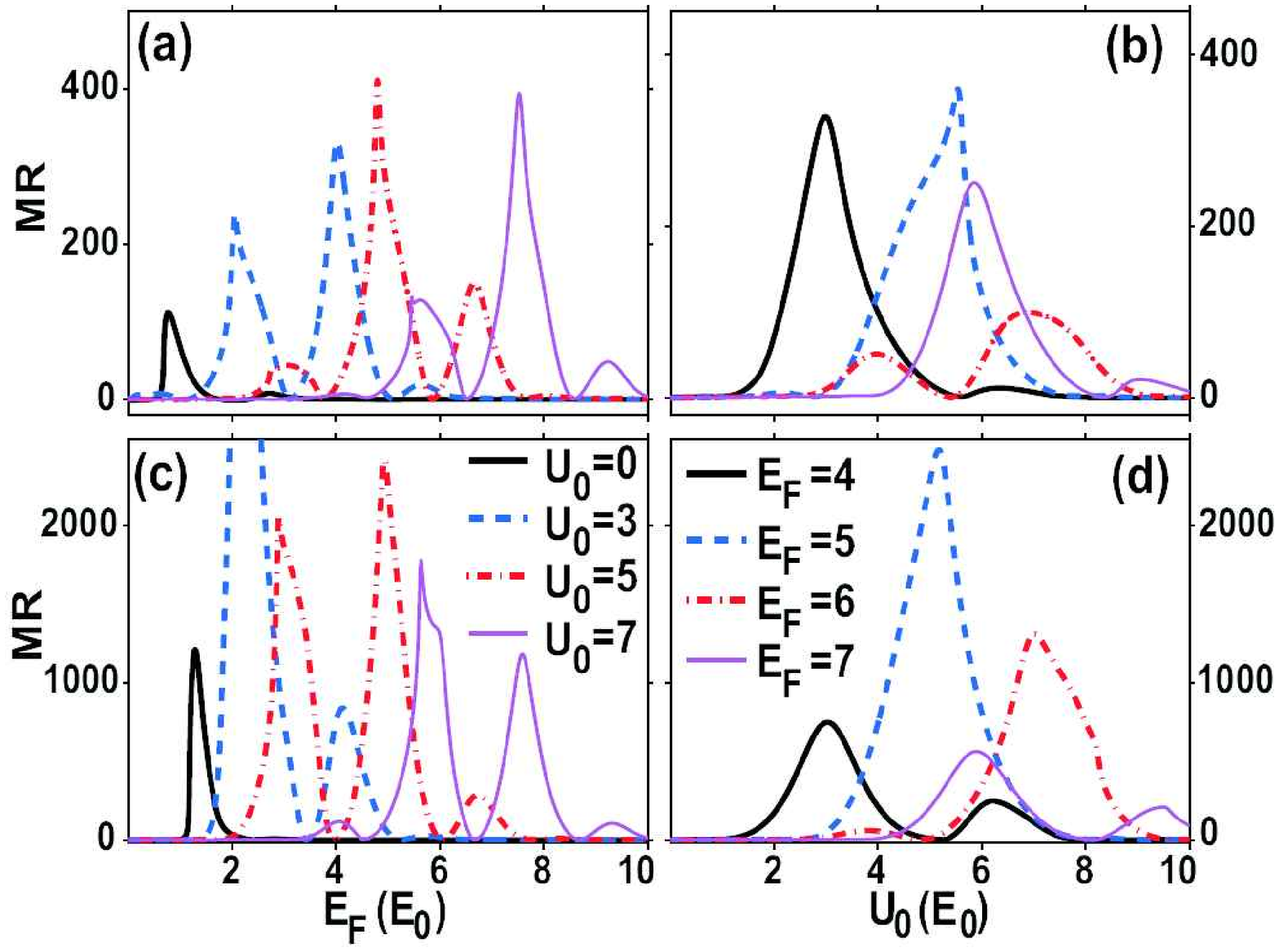}
\caption{(Color online) MR ratio as a function of [(a) and (c)]
Fermi energy or [(b) and (d)] electric barrier height for
electrons traversing the considered magnetic-electric barrier
structure. In (a) and (b) simplified magnetic field profiles are
utilized and the device parameters used in the calculations are
$d=D=1$ and $B=2$. In (c) and (d) realistic magnetic field
profiles are taken. In the calculations we assume that both
ferromagnetic stripes have a rectangular cross section of width
$d=1$ and height $d_z=0.6$ and magnetization
$\protect\mu_0M_x=1.8$ T (for cobalt material), placed at a
distance of $z_0=0.2$ on top of the graphene plane. Their distance
is $D=1$. } \label{MR}
\end{figure}

As demonstrated above, the transmission features for the P and AP
configurations are quite distinct. Such a difference is also exhibited in
the measurable quantity, the conductance $G$. In Fig.~\ref{Conductance} the
conductance is plotted as a function of the Fermi energy for several heights
of the electric barrier. Resonant peaks can be observed in the conductance
spectrum for both P and AP alignments. For the P alignment, the conductance
is finite (larger than 0.1 in most cases in Fig.~\ref{Conductance}). For the
AP alignment, the conductance is almost zero within a broad energy interval
[covering $(U_{0}-B,U_{0}+B)$] except for several sharp conductance peaks.
In this energy region $G_{AP}$ is depleted by the magnetic barrier whereas $%
G_{P}$ is finite. Away from this transmission-blocking region,
$G_{AP}$ essentially increases with the Fermi energy and is
primarily contributed by
the propagating modes. The normalized difference between $G_{P}$ and $G_{AP}$%
, i.e., the TMR ratio $MR=(G_{P}-G_{AP})/G_{AP}$, is presented in Fig.~\ref%
{MR}(a). In the absence of the electric barrier, high values of
$MR$ are located in the low Fermi energy region, as a result of
the strong suppression of transmission in the AP alignment. The
inclusion of an electric barrier shifts the transmission-blocking
region and, thus, can be used to adjust the MR ratio. The latter
is obviously reflected in Fig.~\ref{MR}(b).

In the above analysis, we take simplified magnetic field profiles
to illustrate the operating principles of the proposed device. In
realistic cases the modulated magnetic field $B_{z}(x)$ has the
smoothing variations on the scale of graphene lattice spacing
($a=0.246$ nm). When both FM
stripes have the same rectangular cross section and magnetization along the $x$%
-direction, the generated magnetic field profiles for the P and AP
alignments can be obtained analytically.\cite{profile} For the
parameters given in the figure caption the calculated MR ratio is
shown in Figs.~\ref{MR}(c) and \ref{MR}(d). The calculation shows
that the conductance of the device has a variation similar to that
in Figs.~\ref{MR}(a) and \ref{MR}(b). The TMR ratio remains large
and exhibits rich variations as the electric barrier height
increases. Since ferromagnetic elements with a submicron scale
have been successfully fabricated on top of a two-dimensional
electron system\cite{FM} and
dielectric layers on monolayered graphene have been realized recently,\cite%
{gate control1,gate control2} our considered structure is realizable with
current technology.

In summary, we have investigated the transport features of a
graphene monolayer under the modulation of both a magnetic double
barrier and an electric barrier, where the magnetic double barrier
is provided by depositing two parallel ferromagnetic stripes with
magnetizations along the current direction. The results indicate
that for the AP magnetization configuration the transmission of
electrons in graphene can be drastically suppressed for all
incident angles. When in the P alignment the Klein tunneling can
be generally observed at specific oblique incident directions
rather than the normal incidence. The difference of
wave-vector-dependent transmission for two magnetization
configurations (P/AP) leads to a large TMR ratio, which can be
further adjusted by the electric barrier. Note that different thin
dielectric layers atop graphene sheets have been fabricated and
then the top gates can be formed by means of standard e-beam
lithography.\cite{gate control1,gate control2} The deposition of
ferromagnetic materials on insulating layers has been widely
adopted to create local magnetic field modulations of the
underlying 2D semiconducting systems.\cite{magnetic
barrier1,magnetic barrier2} Thus our proposed device is within the
realizable scope of current technological advances.

F. Zhai was supported by the training fund of young teachers at Dalian
University of Technology (Grant No. 893208) and the NSFC (Grant No.
10704013). K. Chang was supported by the NSFC (Grant No. 60525405)and the
knowledge innovation project of CAS.

\end{document}